\documentclass[aps,prb,amsmath,amssymb,amsfont,showpacs,superscriptaddress,reprint]{revtex4-1}
\usepackage{amsmath}
\usepackage[dvipsnames]{xcolor}
\usepackage[artemisia]{textgreek}  

\usepackage{graphicx}     

\begin{document}

\title{Heating causes non-linear microwave absorption anomaly in single wall carbon nanotubes}

\author{B. G. M\'{a}rkus}
\affiliation{Department of Physics, Budapest University of Technology and Economics and MTA-BME Lend\"{u}let Spintronics Research Group (PROSPIN), P.O. Box 91, H-1521 Budapest, Hungary}

\author{B. Gy\"{u}re-Garami}
\affiliation{Department of Physics, Budapest University of Technology and Economics and MTA-BME Lend\"{u}let Spintronics Research Group (PROSPIN), P.O. Box 91, H-1521 Budapest, Hungary}

\author{O. S\'{a}gi}
\affiliation{Department of Physics, Budapest University of Technology and Economics and MTA-BME Lend\"{u}let Spintronics Research Group (PROSPIN), P.O. Box 91, H-1521 Budapest, Hungary}

\author{G. Cs\H{o}sz}
\affiliation{Department of Physics, Budapest University of Technology and Economics and MTA-BME Lend\"{u}let Spintronics Research Group (PROSPIN), P.O. Box 91, H-1521 Budapest, Hungary}

\author{F. M\'{a}rkus}
\affiliation{Department of Physics, Budapest University of Technology and Economics and MTA-BME Lend\"{u}let Spintronics Research Group (PROSPIN), P.O. Box 91, H-1521 Budapest, Hungary}

\author{F. Simon}
\affiliation{Department of Physics, Budapest University of Technology and Economics and MTA-BME Lend\"{u}let Spintronics Research Group (PROSPIN), P.O. Box 91, H-1521 Budapest, Hungary}


\begin{abstract}Microwave impedance measurements indicate a non-linear absorption anomaly in single wall carbon nanotubes at low temperatures (below $20$ K). We investigate the nature of the anomaly using a time resolved microwave impedance measurement technique. It proves that the anomaly has an extremely slow, a few hundred second long dynamics. This strongly suggests that the anomaly is not caused by an intrinsic electronic effect and that it is rather due to a slow heat exchange between the sample and the environment.\end{abstract}

\maketitle

\section{Introduction}

Conductivity of single-wall carbon nanotubes (SWCNTs) remains an intensively studied subject due to the potential applications of SWCNTs as interconnects \cite{DekkerNAT1997} or switching elements \cite{BachtoldSCI2001}. It is well known that SWCNTs \cite{IijimaNAT1993,BethuneNAT1993} are quasi one-dimensional conductors \cite{HamadaPRL1992} whose electronic properties are determined by the structure. SWCNTs can be either metallic or semiconducting in a $1:2$ ratio of abundance for mean tube diameters above $\sim 1$ nm \cite{DresselhausTubes}. This intermixed nature of conducting properties on one hand are advantageous (as it allows for contacts and semiconducting elements in the same sample) but it is on the other hand a hindrance as sorting according to the metallicity is required.

It was clarified that metallic SWCNTs form the so-called Tomonaga--Luttinger liquid \cite{Tomonaga,Luttinger} phase \cite{EggerPRL1997,KaneBalentsFischerPRL1997,BockrathNAT,KatauraNAT2003,BachtoldPRL2004,PichlerPRL2004,IharaEPL2010} while the semiconducting SWCNTs have a diameter dependent gap of about $0.5-2$ eV which is appropriate for transistor or optoelectronic applications. In addition to their underlying tubular structure (which we call "primary" structure), SWCNTs form a hexagonal arrangement, known as bundles (the "secondary" structure), due to van der Waals force \cite{Thess:Science273:483:(1996)}. The bundles also form a tertiary structure which is better known as an SWCNT "spaghetti". Eventually the conduction is mainly limited by the bundle-bundle interaction in the tertiary structure (e.g. in the form of Schottky barriers) even when in the underlying metallic SWCNTs provide a percolated metallic interconnected network.

Resistivity for a macroscopic SWCNT sample was shown to be characterized by a semiconductor-like temperature dependence \cite{KaiserPRB1998}:
\begin{equation}
\varrho \propto \textrm{e}^{T_\text{c}/T},
\label{resistivity}
\end{equation}
with a $T_\text{c}=10\dots100$ K (using the notation in Ref. \onlinecite{KaiserPRB1998}). This is due to the bundle-bundle contacts that act as transport barriers. Microwave conductivity, based on the microwave cavity perturbation \cite{buravov71,Klein1993,Gruner1993}, is a contactless method with a large sensitivity to relative changes in the conductivity. It proved to be efficient e.g. for the characterization of air sensitive fulleride conductors \cite{BommeliPRB1995,MaedaPRL} and black phosphorus \cite{MarkusPSSb2017}, where conventional contact methods are impractical. The technique relies on the measurement of the cavity quality factor, $Q$, that changes in the presence of the samples. For powder samples, such as fullerides or SWCNTs, the method has the highest sensitivity for the electric conductivity when the sample is placed into the maximum of the microwave magnetic field and the node of the electric one \cite{MaedaPRL}. It turns out that $Q \propto \varrho$, which allows to study the relative variation of the latter quantity. The microwave technique essentially yields the DC conductivity since the microwave frequency ($\sim 10$ GHz) is much lower than the typical value of the plasma frequency (a few $100$ THz).

SWCNTs are considered to be efficient microwave absorbers (that can find applications in e.g. the defense industry) due to their porous nature, excellent thermal stability, and good heat conductivity \cite{HonePRB1999}. Nevertheless, a thorough description of the fundamental microwave absorption properties is necessary prior to their successful applications. 

Corzilius \emph{et al.} \cite{CorziliusPRB2007,CorziliusPSSB2008} reported microwave conductivity measurement on SWCNTs using the cavity perturbation method. A semiconducting-like temperature dependence of $Q\propto \varrho$ was observed above $\sim 20\,\text{K}$. However, an unexpected low temperature anomaly was also observed; the cavity quality factor showed a maximum followed by a decreasing $Q$ on further lowering the temperature. The temperature of the maximum depended on the applied power, its value shifting toward higher temperatures with increasing power. It was argued in Refs. \onlinecite{CorziliusPRB2007,CorziliusPSSB2008} that the anomaly originates from a true electronic effect and a possible occurrence of superconductivity was suggested. The effect was reproduced in Ref. \onlinecite{KarsaPSSB} on different SWCNT samples (arc discharge instead of "super-growth" CVD samples) and using a different detection scheme (fixed frequency irradiation instead of a frequency sweeping in Refs. \onlinecite{CorziliusPRB2007,CorziliusPSSB2008}). This showed that the effect is ubiquitous to SWCNTs. 

Herein, we study this non-linear microwave absorption effect further. The focus is on unraveling the nature of the anomaly. We employ a recently developed time resolved $Q$ measurement \cite{GyureRSI,GyurePreprint} which allows one to monitor the cavity $Q$ factor change with a microsecond time resolution. The data taken at $10$ K indicates that the microwave irradiation induced $Q$ factor change has a very long time constant (a few tens or hundred seconds). For the opposite process, i.e. after irradiating the cavity with microwaves, it recovers to its low microwave power value after a similarly long waiting time. This behavior shows that the non-linear microwave absorption in SWCNTs is a process with an extremely slow dynamics, which strongly argues that it is related to heating effects rather than to an electronic process which should follow a prompt behavior.

\section{Experimental}

\begin{figure}[h!]
\begin{center}
\includegraphics[width=\linewidth]{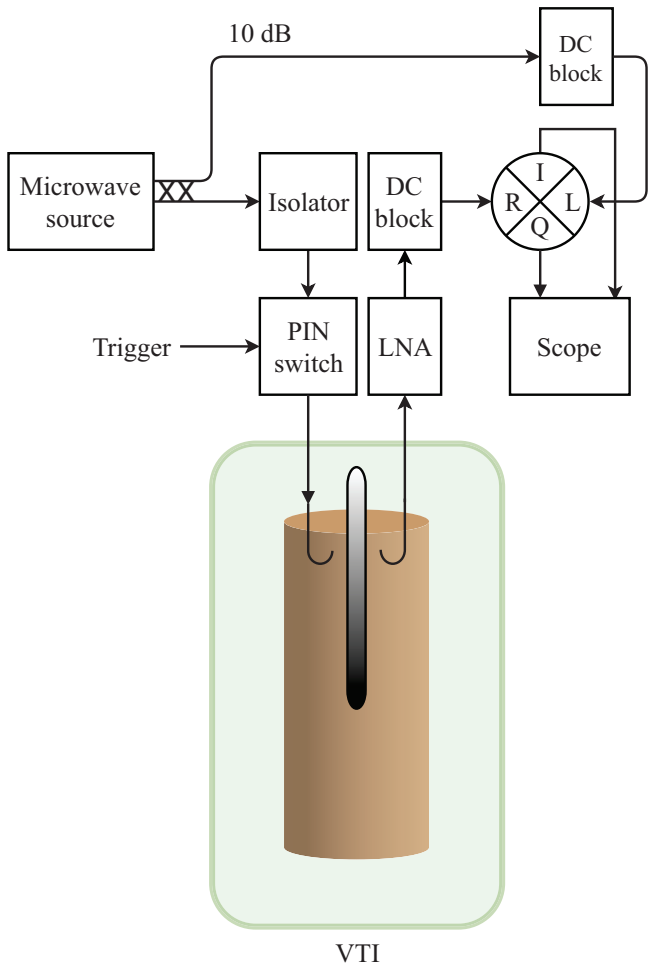}
\caption{Experimental setup used to measure the cavity $Q$ factor with high time resolution. The sample is placed inside a TE011 copper cavity which is inside the VTI. The cavity is coupled to the microwave circuit with two inductive loop antennae in a transmission configuration. A microwave source drives the LO of an IQ mixer and the cavity switch off transient is measured when microwaves are stopped with a PIN diode. A low noise amplifier is also part of the circuit as well as two DC blocks to remove unwanted switching spikes.}
\label{Fig1_ExpSetup}
\end{center}
\end{figure}

We studied SWCNT samples from the same batch as in Ref. \onlinecite{KarsaPSSB} and these samples were used for a number of other studies including Raman \cite{SimonPRB2005,SimonPRB2006}, magnetic resonance \cite{SimonPRL2005,SingerPRL2005,SimonPRL2006}, and EELS studies \cite{SimonCPL2004}. As a result, these are thoroughly characterized samples with well-known diameter distribution, number of tube defects, and electronic properties. In brief, these are commercial arc-discharge grown SWCNTs (Nanocarblab, Russia, Moscow) with mean diameter of $1.4$ nm and diameter standard deviation of $0.1$ nm. A fine powder of the sample (about $5$ mg {\color{black}with a diameter of $3$ mm and height of $2$ mm}) was placed in a quartz tube, first heated to $500~^{\circ}$C to remove contaminations and then sealed under $20$ mbar helium exchange gas for the low temperature measurements. It is important to note that this low pressure helium gas inside the quartz tube remains in the gas form down to $10$ K of our cryostat measurements, it therefore retains the good heat conduction properties. The sample is inside a TE011 copper cavity with an unloaded $Q\approx 10,000$ and resonance frequency of $f_0\approx 11.2$ GHz. We do not correct the measured loaded $Q$ values with the unloaded (empty) $Q$ values since the sample provides much enough load and such a correction would not affect the interpretation of the data. The cavity is inside the VTI of a cryocooled cryostat (Cryogenic Inc.) which allows operation down to $3.2$ K. We performed most measurements at $10$ K, where the cryostat can be well stabilized and efficiently operated for the required long measurement times. 

The cavity is coupled to the rest of the microwave circuit with two loop antennae in a transmission configuration. The transmission is advantageous as it avoids unwanted parasitic microwave signals which are related to the use of microwave duplexers in a reflection geometry \cite{PooleBook}. This microwave setup is detailed in Refs. \onlinecite{GyureRSI,GyurePreprint}. It consists of a source which drives the LO of an IQ mixer as a detection circuit. Microwaves with variable power excite the cavity whose switch off transient is detected with a fast oscilloscope whose signal is Fourier transformed. The latter yields the cavity resonance profiles and a Lorentzian fit gives the resonance bandwidth, $\Delta f$, and the resonance frequency $f_0$, then $Q=f_0/\Delta f$ is obtained. The switch off is generated with a rapid PIN diode and the transmitted microwaves are amplified and filtered with a DC block against the PIN diode switching signal. We used attenuators in order to keep the RF signal level constant on the mixer, independent of the power which irradiates the cavity.

The loaded cavity has a typical $Q \approx 3,000-5,000$ which corresponds to a cavity transient time $\tau=Q/2\pi f_0\approx 50$ ns. This means that the method allows essentially to measure the cavity $Q$ factor down to about $50~\text{ns}-1~\mu\text{s}$ (the value depends on the $Q$ value). We also employ a more conventional frequency swept $Q$ measurement method in order to reproduce the previous results on the non-linear microwave absorption. 

\section{Results}

\begin{figure}[h!]
\begin{center}
\includegraphics[width=\linewidth]{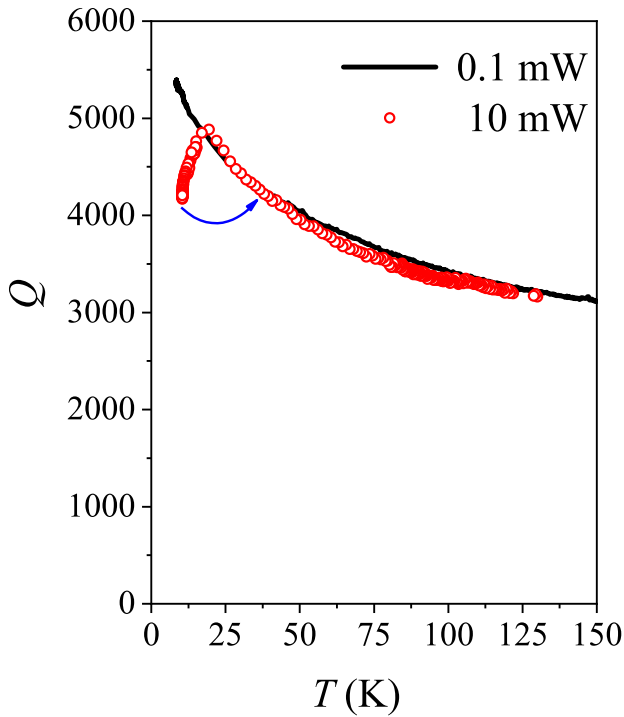}
\caption{Cavity $Q$ factor as a function of temperature for SWCNTs for two different values of the irradiation power. The result reproduces data in Refs. \onlinecite{CorziliusPRB2007,CorziliusPSSB2008,KarsaPSSB}. Note the presence of the non-linear microwave absorption at around $20$ K. The curved arrow points to data points which have the same $Q$ value but are at a different temperature.}
\label{Fig2_Reproduction}
\end{center}
\end{figure}

In Fig. \ref{Fig2_Reproduction}., we show the cavity $Q$ factor as a function of temperature for two different values of the irradiation power. Above $\sim 20\,\text{K}$, the sample behaves as a semiconductor in agreement with the expectation, since $Q\propto \varrho$ in this type of measurement. This is followed down to the lowest temperatures for a low level of microwave irradiation. However for a larger irradiation, an unexpected downturn of the $Q$ factor is observed. This reproduces well the observation in Refs. \onlinecite{CorziliusPRB2007,CorziliusPSSB2008,KarsaPSSB}. It is tempting to associate this downturn to the heating of the sample to higher temperatures, which is schematically indicated by an arrow in the figure. We envisage that if the sample was heated from $10$ K to $30$ K, as the curved arrow in Fig. \ref{Fig2_Reproduction}. hints, the downturning of $Q$ factor effect could be explained. The alternative explanation in Refs. \onlinecite{CorziliusPRB2007,CorziliusPSSB2008} was {\color{black}that} it is a true electronic effect. To settle this issue, we performed time resolved $Q$ factor measurements.

Before presenting our results, we briefly discuss why conventional $Q$ factor measurements are not capable of providing information with a good time resolution. The simplest and most conventional way to obtain $Q$ values of microwave cavities is to sweep the frequency of a source and to detect the cavity response \cite{LuitenReview,KajfezReview}. The frequency sweep time is usually limited to a few $10$ ms or seconds and is normally repeated several times. This means that the sample and cavity is irradiated with the microwaves for relatively long periods of time, therefore no time resolved $Q$ values are available. Certainly, the lowest limit for any time resolved $Q$ measurement is the time constant of the cavity itself, $\tau$, which can be as long as a few seconds for ultra high $Q$ resonators but its value is typically a $\tau\approx 50-500$ ns for a common copper cavity. 

The recently developed cavity transient method fills the gap between the few $10$ ns and few seconds range as it allows to determine the $Q$ factor for as short as $10 \times \tau$. It is based on irradiating the cavity on resonance with microwaves for about $5-10 \times \tau$ in a pulsed manner \cite{GyureRSI,GyurePreprint}. During this time, the cavity accumulates a sizeable energy from the microwave irradiation and stores it. The cavity starts to emit the stored energy immediately following the switch off, whose time decay envelope is an exponential function with the time constant $\tau$. The power of the Fourier transformed signal (considering the real and imaginary outputs of the IQ mixer) yields directly the resonance curve. A fit to the curve yields the $Q$ factor.

\begin{figure}[h!]
\begin{center}
\includegraphics[width=\linewidth]{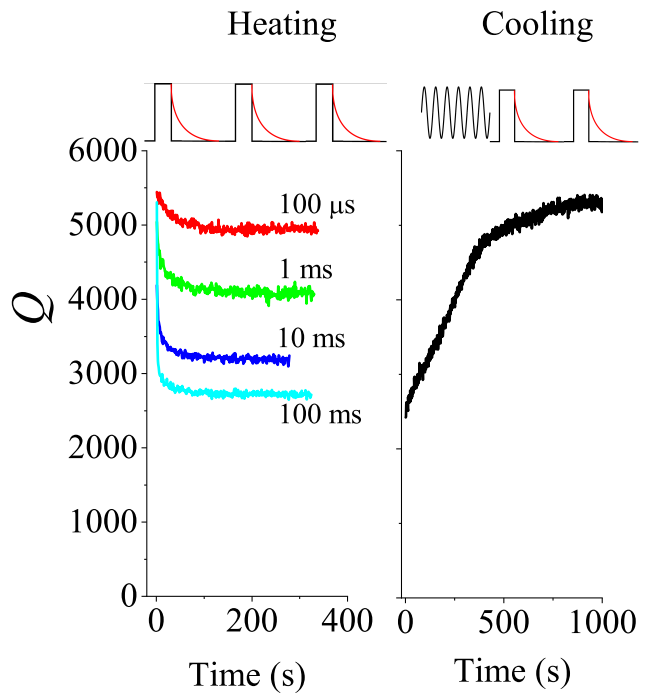}
\caption{Time resolved heating and cooling curves of the cavity $Q$ factor in SWCNTs when the VTI is kept at $10$ K. For heating, the pulse length is varied and all curves start from the same initial $Q$ value. Note the much longer cooling dynamics and also the different horizontal scale for the two types of experiments.}
\label{Fig3_Heating_Cooling}
\end{center}
\end{figure}

To investigate the time dynamics of the non-linear microwave absorption, we performed two types of experiments. In the first type, we start irradiating the sample with short microwave pulses after it was allowed to thermalize for a longer period of time ($200\,\text{s}$) without microwave irradiation. The pulses themselves serve two purposes: they irradiate the sample in a controllable manner and also the pulses allow to read out the state of the cavity. The second experiment consists of irradiating the sample for a long period of time ($200$ seconds) and then applying short microwave pulses to read out the $Q$ factor of the cavity. To simplify to subsequent discussion, we refer to these experiments as "heating" and "cooling". 

In Fig. \ref{Fig3_Heating_Cooling}., we show the variation of the cavity $Q$ factor during heating and cooling in a time resolved manner. For both types of measurements, we detect switch off cavity transients after the microwave pulses with a repetition time of $1$ sec. The dynamics of the system showed that this is sufficient, although repetitions with a much shorter timescale would be possible. The peak power in each pulse was 10 mW, which produces the microwave absorption anomaly according to Fig. \ref{Fig2_Reproduction}. For heating, we employed irradiation pulse lengths of $100~\mu\text{s}$, $1~\text{ms}$, $10~\text{ms}$, and $100~\text{ms}$. We observe that the cavity $Q$ progressively decreases for a given pulse length experiment with the same time constant of about a $100$ sec. The asymptotic $Q$ value is smaller for longer pulses. 

For the cooling experiment, we let the sample thermalize for $200$ sec, when no microwave irradiation is employed. This is followed by a massive irradiation of $10$ mW power applied for $200$ seconds, afterwards read-out pulses with a duration of $1~\mu\text{s}$ were used. The result is also {\color{black}shown} in Fig. \ref{Fig3_Heating_Cooling}. The cooling has a slower dynamics as compared to the heating.

Note that the effect is by no means related to the copper cavity as for other type of samples, e.g. the K$_3$C$_{60}$, we did not observe a similar effect down to the lowest temperatures \cite{Csosz2018}. The observed time dependence of the microwave cavity $Q$ factor change therefore {\color{black}suggests} that it is intrinsic to SWCNTs. In addition, the slow dynamics strongly suggests that it is related to heating or cooling effects as we do not know of another {\color{black}process} which could explain this behavior. In view of this, the previous hint in Fig. \ref{Fig2_Reproduction}., i.e. that the downturn of $Q$ with increasing power is due to heating, is reinforced.

\section{Discussion}
The cause of the anomalous heating effect present in single walled nanotube bundles is probably related to its unique heat conduction properties, which is expected to be very high, according to theoretical predictions around $2000-6000$ W$/$m$\cdot$K \cite{Che2000,Grujicic2004,Han2011}. However these extremely high values are only reported for along the {\color{black}nanotube} axis \cite{YuNanoLett2005} {\color{black} at room temperature}. In addition, the heat conductivity vanishes rapidly with decreasing temperature and drops below $2000$ W$/$m$\cdot$K at $100$ K, and the process is generated by {\color{black}acoustic phonon freeze-out} \cite{HonePRB1999}. 

The thermal conductivity is expected to actually become $0$ as temperature approaches zero. The situation worsens when the investigated sample is bulk, which {\color{black}causes} the drop of {\color{black}$\kappa_{\parallel}$} to the range of $50$ to $200$ W$/$m$\cdot$K at room temperature \cite{Hone2000,HoneAPA2002,Fischer2003}. However, these data were also taken for magnetic field aligned SWCNTs (and measurements were done in the {\color{black}nanotube} axis). In contrast, perpendicular to this axis the values of {\color{black}$\kappa_{\perp}$} are at least $3$ orders of magnitude smaller, around $1.5$ W$/$m$\cdot$K {\color{black}at 300 K \cite{SinhaNanopart2005}. The thermal conductivity is expected to be at least 10 times smaller than this value at 10 K due to the phonon freeze-out \cite{HonePRB1999}}. 

Taking into account that our sample consists of unaligned tubes with short lengths, one expects the thermal conductivity to be as low as in glasses at low temperatures. Not to mention that shorter length, lattice defects and vacancies also lower the value of {\color{black}$\kappa$, each by a factor of $1-3$ \cite{Che2000}}. {\color{black}Besides the thermal conductivity, thermal diffusivity may also play a role in the observed phenomenon.} Considering these findings and Fourier's law of thermal conduction {\color{black} with homogeneous heating}, it can occur in a mm sized sample (such as our samples) that even when the tubes at the cryostat interface are at $10$ K, the tubes inside the bulk of the sample can be at $100$ K or even higher temperatures. This cannot be compensated even with the surrounding He exchange gas. Heating of the inside of the sample occurs as the porous powder does not limit the penetration of microwaves. The heating forms \emph{hot spots} and during the $Q$ measurements, these are averaged together with the colder SWCNTs on the outside of the sample, thus an effective drop in the resistivity is observed.

\section{Summary}

We presented a time resolved measurement of microwave conductivity in single wall carbon nanotubes. The experiments indicate a non-linear microwave absorption anomaly which occurs with an extremely slow timescale ({\color{black}of} a few hundred seconds). This suggest that heating of the SWCNTs is responsible for the anomalous behavior as a truly electronic effect is expected to occur on a much faster timescale. The result may explain some of the low temperature microwave anomalies in carbon nanotubes. 

\section{Acknowledgement}
Work supported by the Hungarian National Research, Development and Innovation Office (NKFIH) Grant Nr. K119442 and 2017-1.2.1-NKP-2017-00001.

\bibliography{Tubes2012June}

\begin{thebibliography}{47}%
\makeatletter
\providecommand \@ifxundefined [1]{%
 \@ifx{#1\undefined}
}%
\providecommand \@ifnum [1]{%
 \ifnum #1\expandafter \@firstoftwo
 \else \expandafter \@secondoftwo
 \fi
}%
\providecommand \@ifx [1]{%
 \ifx #1\expandafter \@firstoftwo
 \else \expandafter \@secondoftwo
 \fi
}%
\providecommand \natexlab [1]{#1}%
\providecommand \enquote  [1]{``#1''}%
\providecommand \bibnamefont  [1]{#1}%
\providecommand \bibfnamefont [1]{#1}%
\providecommand \citenamefont [1]{#1}%
\providecommand \href@noop [0]{\@secondoftwo}%
\providecommand \href [0]{\begingroup \@sanitize@url \@href}%
\providecommand \@href[1]{\@@startlink{#1}\@@href}%
\providecommand \@@href[1]{\endgroup#1\@@endlink}%
\providecommand \@sanitize@url [0]{\catcode `\\12\catcode `\$12\catcode
  `\&12\catcode `\#12\catcode `\^12\catcode `\_12\catcode `\%12\relax}%
\providecommand \@@startlink[1]{}%
\providecommand \@@endlink[0]{}%
\providecommand \url  [0]{\begingroup\@sanitize@url \@url }%
\providecommand \@url [1]{\endgroup\@href {#1}{\urlprefix }}%
\providecommand \urlprefix  [0]{URL }%
\providecommand \Eprint [0]{\href }%
\providecommand \doibase [0]{http://dx.doi.org/}%
\providecommand \selectlanguage [0]{\@gobble}%
\providecommand \bibinfo  [0]{\@secondoftwo}%
\providecommand \bibfield  [0]{\@secondoftwo}%
\providecommand \translation [1]{[#1]}%
\providecommand \BibitemOpen [0]{}%
\providecommand \bibitemStop [0]{}%
\providecommand \bibitemNoStop [0]{.\EOS\space}%
\providecommand \EOS [0]{\spacefactor3000\relax}%
\providecommand \BibitemShut  [1]{\csname bibitem#1\endcsname}%
\let\auto@bib@innerbib\@empty
\bibitem [{\citenamefont {Tans}\ \emph {et~al.}(1997)\citenamefont {Tans},
  \citenamefont {Devoret}, \citenamefont {Dai}, \citenamefont {Thess},
  \citenamefont {Smalley}, \citenamefont {Geerligs},\ and\ \citenamefont
  {Dekker}}]{DekkerNAT1997}%
  \BibitemOpen
  \bibfield  {author} {\bibinfo {author} {\bibfnamefont {S.~J.}\ \bibnamefont
  {Tans}}, \bibinfo {author} {\bibfnamefont {M.~H.}\ \bibnamefont {Devoret}},
  \bibinfo {author} {\bibfnamefont {H.}~\bibnamefont {Dai}}, \bibinfo {author}
  {\bibfnamefont {A.}~\bibnamefont {Thess}}, \bibinfo {author} {\bibfnamefont
  {R.~E.}\ \bibnamefont {Smalley}}, \bibinfo {author} {\bibfnamefont {L.~J.}\
  \bibnamefont {Geerligs}}, \ and\ \bibinfo {author} {\bibfnamefont
  {C.}~\bibnamefont {Dekker}},\ }\href@noop {} {\bibfield  {journal} {\bibinfo
  {journal} {Nature}\ }\textbf {\bibinfo {volume} {386}},\ \bibinfo {pages}
  {474 } (\bibinfo {year} {1997})}\BibitemShut {NoStop}%
\bibitem [{\citenamefont {Bachtold}\ \emph {et~al.}(2001)\citenamefont
  {Bachtold}, \citenamefont {Hadley}, \citenamefont {Nakanishi},\ and\
  \citenamefont {Dekker}}]{BachtoldSCI2001}%
  \BibitemOpen
  \bibfield  {author} {\bibinfo {author} {\bibfnamefont {A.}~\bibnamefont
  {Bachtold}}, \bibinfo {author} {\bibfnamefont {P.}~\bibnamefont {Hadley}},
  \bibinfo {author} {\bibfnamefont {T.}~\bibnamefont {Nakanishi}}, \ and\
  \bibinfo {author} {\bibfnamefont {C.}~\bibnamefont {Dekker}},\ }\href@noop {}
  {\bibfield  {journal} {\bibinfo  {journal} {Science}\ }\textbf {\bibinfo
  {volume} {294}},\ \bibinfo {pages} {1317} (\bibinfo {year}
  {2001})}\BibitemShut {NoStop}%
\bibitem [{\citenamefont {Iijima}\ and\ \citenamefont
  {Ichihashi}(1993)}]{IijimaNAT1993}%
  \BibitemOpen
  \bibfield  {author} {\bibinfo {author} {\bibfnamefont {S.}~\bibnamefont
  {Iijima}}\ and\ \bibinfo {author} {\bibfnamefont {T.}~\bibnamefont
  {Ichihashi}},\ }\href@noop {} {\bibfield  {journal} {\bibinfo  {journal}
  {Nature}\ }\textbf {\bibinfo {volume} {363}},\ \bibinfo {pages} {603}
  (\bibinfo {year} {1993})}\BibitemShut {NoStop}%
\bibitem [{\citenamefont {Bethune}\ \emph {et~al.}(1993)\citenamefont
  {Bethune}, \citenamefont {Kiang}, \citenamefont {DeVries}, \citenamefont
  {Gorman}, \citenamefont {Savoy},\ and\ \citenamefont
  {Beyers}}]{BethuneNAT1993}%
  \BibitemOpen
  \bibfield  {author} {\bibinfo {author} {\bibfnamefont {D.~S.}\ \bibnamefont
  {Bethune}}, \bibinfo {author} {\bibfnamefont {C.~H.}\ \bibnamefont {Kiang}},
  \bibinfo {author} {\bibfnamefont {M.~S.}\ \bibnamefont {DeVries}}, \bibinfo
  {author} {\bibfnamefont {G.}~\bibnamefont {Gorman}}, \bibinfo {author}
  {\bibfnamefont {R.}~\bibnamefont {Savoy}}, \ and\ \bibinfo {author}
  {\bibfnamefont {R.}~\bibnamefont {Beyers}},\ }\href@noop {} {\bibfield
  {journal} {\bibinfo  {journal} {Nature}\ }\textbf {\bibinfo {volume} {363}},\
  \bibinfo {pages} {605} (\bibinfo {year} {1993})}\BibitemShut {NoStop}%
\bibitem [{\citenamefont {Hamada}\ \emph {et~al.}(1992)\citenamefont {Hamada},
  \citenamefont {Sawada},\ and\ \citenamefont {Oshiyama}}]{HamadaPRL1992}%
  \BibitemOpen
  \bibfield  {author} {\bibinfo {author} {\bibfnamefont {N.}~\bibnamefont
  {Hamada}}, \bibinfo {author} {\bibfnamefont {S.}~\bibnamefont {Sawada}}, \
  and\ \bibinfo {author} {\bibfnamefont {A.}~\bibnamefont {Oshiyama}},\
  }\href@noop {} {\bibfield  {journal} {\bibinfo  {journal} {Phys. Rev. Lett.}\
  }\textbf {\bibinfo {volume} {68}},\ \bibinfo {pages} {1579} (\bibinfo {year}
  {1992})}\BibitemShut {NoStop}%
\bibitem [{\citenamefont {Saito}\ \emph {et~al.}(1998)\citenamefont {Saito},
  \citenamefont {Dresselhaus},\ and\ \citenamefont
  {Dresselhaus}}]{DresselhausTubes}%
  \BibitemOpen
  \bibfield  {author} {\bibinfo {author} {\bibfnamefont {R.}~\bibnamefont
  {Saito}}, \bibinfo {author} {\bibfnamefont {G.}~\bibnamefont {Dresselhaus}},
  \ and\ \bibinfo {author} {\bibfnamefont {M.}~\bibnamefont {Dresselhaus}},\
  }\href@noop {} {\emph {\bibinfo {title} {Physical Properties of Carbon
  Nanotubes}}}\ (\bibinfo  {publisher} {Imperial College Press},\ \bibinfo
  {year} {1998})\BibitemShut {NoStop}%
\bibitem [{\citenamefont {Tomonaga}(1950)}]{Tomonaga}%
  \BibitemOpen
  \bibfield  {author} {\bibinfo {author} {\bibfnamefont {S.}~\bibnamefont
  {Tomonaga}},\ }\href@noop {} {\bibfield  {journal} {\bibinfo  {journal}
  {Prog. Theor. Phys.}\ }\textbf {\bibinfo {volume} {5}},\ \bibinfo {pages}
  {349} (\bibinfo {year} {1950})}\BibitemShut {NoStop}%
\bibitem [{\citenamefont {Luttinger}(1963)}]{Luttinger}%
  \BibitemOpen
  \bibfield  {author} {\bibinfo {author} {\bibfnamefont {J.~M.}\ \bibnamefont
  {Luttinger}},\ }\href@noop {} {\bibfield  {journal} {\bibinfo  {journal} {J.
  Math. Phys.}\ }\textbf {\bibinfo {volume} {4}},\ \bibinfo {pages} {1154}
  (\bibinfo {year} {1963})}\BibitemShut {NoStop}%
\bibitem [{\citenamefont {Egger}\ and\ \citenamefont
  {Gogolin}(1997)}]{EggerPRL1997}%
  \BibitemOpen
  \bibfield  {author} {\bibinfo {author} {\bibfnamefont {R.}~\bibnamefont
  {Egger}}\ and\ \bibinfo {author} {\bibfnamefont {A.~O.}\ \bibnamefont
  {Gogolin}},\ }\href@noop {} {\bibfield  {journal} {\bibinfo  {journal} {Phys.
  Rev. Lett.}\ }\textbf {\bibinfo {volume} {79}},\ \bibinfo {pages} {5082}
  (\bibinfo {year} {1997})}\BibitemShut {NoStop}%
\bibitem [{\citenamefont {Kane}\ \emph {et~al.}(1997)\citenamefont {Kane},
  \citenamefont {Balents},\ and\ \citenamefont
  {Fisher}}]{KaneBalentsFischerPRL1997}%
  \BibitemOpen
  \bibfield  {author} {\bibinfo {author} {\bibfnamefont {C.}~\bibnamefont
  {Kane}}, \bibinfo {author} {\bibfnamefont {L.}~\bibnamefont {Balents}}, \
  and\ \bibinfo {author} {\bibfnamefont {M.~P.~A.}\ \bibnamefont {Fisher}},\
  }\href@noop {} {\bibfield  {journal} {\bibinfo  {journal} {Phys. Rev. Lett.}\
  }\textbf {\bibinfo {volume} {79}},\ \bibinfo {pages} {5086} (\bibinfo {year}
  {1997})}\BibitemShut {NoStop}%
\bibitem [{\citenamefont {Bockrath}\ \emph {et~al.}(1999)\citenamefont
  {Bockrath}, \citenamefont {Cobden}, \citenamefont {Lu}, \citenamefont
  {Rinzler}, \citenamefont {Smalley}, \citenamefont {Balents},\ and\
  \citenamefont {McEuen}}]{BockrathNAT}%
  \BibitemOpen
  \bibfield  {author} {\bibinfo {author} {\bibfnamefont {M.}~\bibnamefont
  {Bockrath}}, \bibinfo {author} {\bibfnamefont {D.~H.}\ \bibnamefont
  {Cobden}}, \bibinfo {author} {\bibfnamefont {J.}~\bibnamefont {Lu}}, \bibinfo
  {author} {\bibfnamefont {A.~G.}\ \bibnamefont {Rinzler}}, \bibinfo {author}
  {\bibfnamefont {R.~E.}\ \bibnamefont {Smalley}}, \bibinfo {author}
  {\bibfnamefont {L.}~\bibnamefont {Balents}}, \ and\ \bibinfo {author}
  {\bibfnamefont {P.~L.}\ \bibnamefont {McEuen}},\ }\href@noop {} {\bibfield
  {journal} {\bibinfo  {journal} {Nature}\ }\textbf {\bibinfo {volume} {397}},\
  \bibinfo {pages} {598 } (\bibinfo {year} {1999})}\BibitemShut {NoStop}%
\bibitem [{\citenamefont {Ishii}\ \emph {et~al.}(2003)\citenamefont {Ishii},
  \citenamefont {Kataura}, \citenamefont {Shiozawa}, \citenamefont {Yoshioka},
  \citenamefont {Otsubo}, \citenamefont {Takayama}, \citenamefont {Miyahara},
  \citenamefont {Suzuki}, \citenamefont {Achiba}, \citenamefont {Nakatake},
  \citenamefont {Narimura}, \citenamefont {Higashiguchi}, \citenamefont
  {Shimada}, \citenamefont {Namatame},\ and\ \citenamefont
  {Taniguchi}}]{KatauraNAT2003}%
  \BibitemOpen
  \bibfield  {author} {\bibinfo {author} {\bibfnamefont {H.}~\bibnamefont
  {Ishii}}, \bibinfo {author} {\bibfnamefont {H.}~\bibnamefont {Kataura}},
  \bibinfo {author} {\bibfnamefont {H.}~\bibnamefont {Shiozawa}}, \bibinfo
  {author} {\bibfnamefont {H.}~\bibnamefont {Yoshioka}}, \bibinfo {author}
  {\bibfnamefont {H.}~\bibnamefont {Otsubo}}, \bibinfo {author} {\bibfnamefont
  {Y.}~\bibnamefont {Takayama}}, \bibinfo {author} {\bibfnamefont
  {T.}~\bibnamefont {Miyahara}}, \bibinfo {author} {\bibfnamefont
  {S.}~\bibnamefont {Suzuki}}, \bibinfo {author} {\bibfnamefont
  {Y.}~\bibnamefont {Achiba}}, \bibinfo {author} {\bibfnamefont
  {M.}~\bibnamefont {Nakatake}}, \bibinfo {author} {\bibfnamefont
  {T.}~\bibnamefont {Narimura}}, \bibinfo {author} {\bibfnamefont
  {M.}~\bibnamefont {Higashiguchi}}, \bibinfo {author} {\bibfnamefont
  {K.}~\bibnamefont {Shimada}}, \bibinfo {author} {\bibfnamefont
  {H.}~\bibnamefont {Namatame}}, \ and\ \bibinfo {author} {\bibfnamefont
  {M.}~\bibnamefont {Taniguchi}},\ }\href@noop {} {\bibfield  {journal}
  {\bibinfo  {journal} {Nature}\ }\textbf {\bibinfo {volume} {426}},\ \bibinfo
  {pages} {540} (\bibinfo {year} {2003})}\BibitemShut {NoStop}%
\bibitem [{\citenamefont {Gao}\ \emph {et~al.}(2004)\citenamefont {Gao},
  \citenamefont {Komnik}, \citenamefont {Egger}, \citenamefont {Glattli},\ and\
  \citenamefont {Bachtold}}]{BachtoldPRL2004}%
  \BibitemOpen
  \bibfield  {author} {\bibinfo {author} {\bibfnamefont {B.}~\bibnamefont
  {Gao}}, \bibinfo {author} {\bibfnamefont {A.}~\bibnamefont {Komnik}},
  \bibinfo {author} {\bibfnamefont {R.}~\bibnamefont {Egger}}, \bibinfo
  {author} {\bibfnamefont {D.~C.}\ \bibnamefont {Glattli}}, \ and\ \bibinfo
  {author} {\bibfnamefont {C.}~\bibnamefont {Bachtold}},\ }\href@noop {}
  {\bibfield  {journal} {\bibinfo  {journal} {Phys. Rev. Lett.}\ }\textbf
  {\bibinfo {volume} {92}},\ \bibinfo {pages} {216804} (\bibinfo {year}
  {2004})}\BibitemShut {NoStop}%
\bibitem [{\citenamefont {Rauf}\ \emph {et~al.}(2004)\citenamefont {Rauf},
  \citenamefont {Pichler}, \citenamefont {Knupfer}, \citenamefont {Fink},\ and\
  \citenamefont {Kataura}}]{PichlerPRL2004}%
  \BibitemOpen
  \bibfield  {author} {\bibinfo {author} {\bibfnamefont {H.}~\bibnamefont
  {Rauf}}, \bibinfo {author} {\bibfnamefont {T.}~\bibnamefont {Pichler}},
  \bibinfo {author} {\bibfnamefont {M.}~\bibnamefont {Knupfer}}, \bibinfo
  {author} {\bibfnamefont {J.}~\bibnamefont {Fink}}, \ and\ \bibinfo {author}
  {\bibfnamefont {H.}~\bibnamefont {Kataura}},\ }\href@noop {} {\bibfield
  {journal} {\bibinfo  {journal} {Phys. Rev. Lett.}\ }\textbf {\bibinfo
  {volume} {93}},\ \bibinfo {pages} {096805} (\bibinfo {year}
  {2004})}\BibitemShut {NoStop}%
\bibitem [{\citenamefont {Ihara}\ \emph {et~al.}(2010)\citenamefont {Ihara},
  \citenamefont {Wzietek}, \citenamefont {Alloul}, \citenamefont {R\"{u}mmeli},
  \citenamefont {Pichler},\ and\ \citenamefont {Simon}}]{IharaEPL2010}%
  \BibitemOpen
  \bibfield  {author} {\bibinfo {author} {\bibfnamefont {Y.}~\bibnamefont
  {Ihara}}, \bibinfo {author} {\bibfnamefont {P.}~\bibnamefont {Wzietek}},
  \bibinfo {author} {\bibfnamefont {H.}~\bibnamefont {Alloul}}, \bibinfo
  {author} {\bibfnamefont {M.~H.}\ \bibnamefont {R\"{u}mmeli}}, \bibinfo
  {author} {\bibfnamefont {T.}~\bibnamefont {Pichler}}, \ and\ \bibinfo
  {author} {\bibfnamefont {F.}~\bibnamefont {Simon}},\ }\href@noop {}
  {\bibfield  {journal} {\bibinfo  {journal} {Eur. Phys. Lett.}\ }\textbf
  {\bibinfo {volume} {90}},\ \bibinfo {pages} {17004} (\bibinfo {year}
  {2010})}\BibitemShut {NoStop}%
\bibitem [{\citenamefont {Thess}\ \emph {et~al.}(1996)\citenamefont {Thess},
  \citenamefont {Lee}, \citenamefont {Nikolaev}, \citenamefont {Dai},
  \citenamefont {Petit}, \citenamefont {Robert}, \citenamefont {Xu},
  \citenamefont {Lee}, \citenamefont {Kim}, \citenamefont {Rinzler},
  \citenamefont {Colbert}, \citenamefont {Scuseria}, \citenamefont
  {Tom{\'a}nek}, \citenamefont {Fischer},\ and\ \citenamefont
  {Smalley}}]{Thess:Science273:483:(1996)}%
  \BibitemOpen
  \bibfield  {author} {\bibinfo {author} {\bibfnamefont {A.}~\bibnamefont
  {Thess}}, \bibinfo {author} {\bibfnamefont {R.}~\bibnamefont {Lee}}, \bibinfo
  {author} {\bibfnamefont {P.}~\bibnamefont {Nikolaev}}, \bibinfo {author}
  {\bibfnamefont {H.}~\bibnamefont {Dai}}, \bibinfo {author} {\bibfnamefont
  {P.}~\bibnamefont {Petit}}, \bibinfo {author} {\bibfnamefont
  {J.}~\bibnamefont {Robert}}, \bibinfo {author} {\bibfnamefont
  {C.}~\bibnamefont {Xu}}, \bibinfo {author} {\bibfnamefont {Y.~H.}\
  \bibnamefont {Lee}}, \bibinfo {author} {\bibfnamefont {S.~G.}\ \bibnamefont
  {Kim}}, \bibinfo {author} {\bibfnamefont {A.~G.}\ \bibnamefont {Rinzler}},
  \bibinfo {author} {\bibfnamefont {D.~T.}\ \bibnamefont {Colbert}}, \bibinfo
  {author} {\bibfnamefont {G.~E.}\ \bibnamefont {Scuseria}}, \bibinfo {author}
  {\bibfnamefont {D.}~\bibnamefont {Tom{\'a}nek}}, \bibinfo {author}
  {\bibfnamefont {J.~E.}\ \bibnamefont {Fischer}}, \ and\ \bibinfo {author}
  {\bibfnamefont {R.~E.}\ \bibnamefont {Smalley}},\ }\href@noop {} {\bibfield
  {journal} {\bibinfo  {journal} {Science}\ }\textbf {\bibinfo {volume}
  {273}},\ \bibinfo {pages} {483} (\bibinfo {year} {1996})}\BibitemShut
  {NoStop}%
\bibitem [{\citenamefont {Kaiser}\ \emph {et~al.}(1998)\citenamefont {Kaiser},
  \citenamefont {D\"usberg},\ and\ \citenamefont {Roth}}]{KaiserPRB1998}%
  \BibitemOpen
  \bibfield  {author} {\bibinfo {author} {\bibfnamefont {A.~B.}\ \bibnamefont
  {Kaiser}}, \bibinfo {author} {\bibfnamefont {G.}~\bibnamefont {D\"usberg}}, \
  and\ \bibinfo {author} {\bibfnamefont {S.}~\bibnamefont {Roth}},\ }\href@noop
  {} {\bibfield  {journal} {\bibinfo  {journal} {Phys. Rev. B}\ }\textbf
  {\bibinfo {volume} {57}},\ \bibinfo {pages} {1418} (\bibinfo {year}
  {1998})}\BibitemShut {NoStop}%
\bibitem [{\citenamefont {Buravov}\ and\ \citenamefont
  {Shchegolev}(1971)}]{buravov71}%
  \BibitemOpen
  \bibfield  {author} {\bibinfo {author} {\bibfnamefont {L.~I.}\ \bibnamefont
  {Buravov}}\ and\ \bibinfo {author} {\bibfnamefont {I.~F.}\ \bibnamefont
  {Shchegolev}},\ }\href@noop {} {\bibfield  {journal} {\bibinfo  {journal}
  {Instrum.\ Exp.\ Tech.}\ }\textbf {\bibinfo {volume} {14}},\ \bibinfo {pages}
  {528} (\bibinfo {year} {1971})}\BibitemShut {NoStop}%
\bibitem [{\citenamefont {Klein}\ \emph {et~al.}(1993)\citenamefont {Klein},
  \citenamefont {Donovan}, \citenamefont {Dressel},\ and\ \citenamefont
  {Gr{\"u}ner}}]{Klein1993}%
  \BibitemOpen
  \bibfield  {author} {\bibinfo {author} {\bibfnamefont {O.}~\bibnamefont
  {Klein}}, \bibinfo {author} {\bibfnamefont {S.}~\bibnamefont {Donovan}},
  \bibinfo {author} {\bibfnamefont {M.}~\bibnamefont {Dressel}}, \ and\
  \bibinfo {author} {\bibfnamefont {G.}~\bibnamefont {Gr{\"u}ner}},\
  }\href@noop {} {\bibfield  {journal} {\bibinfo  {journal} {Int. J. Infr.
  Mill. Wav.}\ }\textbf {\bibinfo {volume} {14}},\ \bibinfo {pages} {2423}
  (\bibinfo {year} {1993})}\BibitemShut {NoStop}%
\bibitem [{\citenamefont {Donovan}\ \emph {et~al.}(1993)\citenamefont
  {Donovan}, \citenamefont {Klein}, \citenamefont {Dressel}, \citenamefont
  {Holczer},\ and\ \citenamefont {Gr{\"u}ner}}]{Gruner1993}%
  \BibitemOpen
  \bibfield  {author} {\bibinfo {author} {\bibfnamefont {S.}~\bibnamefont
  {Donovan}}, \bibinfo {author} {\bibfnamefont {O.}~\bibnamefont {Klein}},
  \bibinfo {author} {\bibfnamefont {M.}~\bibnamefont {Dressel}}, \bibinfo
  {author} {\bibfnamefont {K.}~\bibnamefont {Holczer}}, \ and\ \bibinfo
  {author} {\bibfnamefont {G.}~\bibnamefont {Gr{\"u}ner}},\ }\href@noop {}
  {\bibfield  {journal} {\bibinfo  {journal} {Int. J. Infr. Mill. Wav.}\
  }\textbf {\bibinfo {volume} {14}},\ \bibinfo {pages} {2459} (\bibinfo {year}
  {1993})}\BibitemShut {NoStop}%
\bibitem [{\citenamefont {Bommeli}\ \emph {et~al.}(1995)\citenamefont
  {Bommeli}, \citenamefont {Degiorgi}, \citenamefont {Wachter}, \citenamefont
  {Legeza}, \citenamefont {J\'anossy}, \citenamefont {Oszl\'anyi},
  \citenamefont {Chauvet},\ and\ \citenamefont {Forr\'o}}]{BommeliPRB1995}%
  \BibitemOpen
  \bibfield  {author} {\bibinfo {author} {\bibfnamefont {F.}~\bibnamefont
  {Bommeli}}, \bibinfo {author} {\bibfnamefont {L.}~\bibnamefont {Degiorgi}},
  \bibinfo {author} {\bibfnamefont {P.}~\bibnamefont {Wachter}}, \bibinfo
  {author} {\bibfnamefont {{\"{O}}.}~\bibnamefont {Legeza}}, \bibinfo {author}
  {\bibfnamefont {A.}~\bibnamefont {J\'anossy}}, \bibinfo {author}
  {\bibfnamefont {G.}~\bibnamefont {Oszl\'anyi}}, \bibinfo {author}
  {\bibfnamefont {O.}~\bibnamefont {Chauvet}}, \ and\ \bibinfo {author}
  {\bibfnamefont {L.}~\bibnamefont {Forr\'o}},\ }\href@noop {} {\bibfield
  {journal} {\bibinfo  {journal} {Phys. Rev. B}\ }\textbf {\bibinfo {volume}
  {51}},\ \bibinfo {pages} {14794} (\bibinfo {year} {1995})}\BibitemShut
  {NoStop}%
\bibitem [{\citenamefont {Kitano}\ \emph {et~al.}(2002)\citenamefont {Kitano},
  \citenamefont {Matsuo}, \citenamefont {Miwa}, \citenamefont {Maeda},
  \citenamefont {Takenobu}, \citenamefont {Iwasa},\ and\ \citenamefont
  {Mitani}}]{MaedaPRL}%
  \BibitemOpen
  \bibfield  {author} {\bibinfo {author} {\bibfnamefont {H.}~\bibnamefont
  {Kitano}}, \bibinfo {author} {\bibfnamefont {R.}~\bibnamefont {Matsuo}},
  \bibinfo {author} {\bibfnamefont {K.}~\bibnamefont {Miwa}}, \bibinfo {author}
  {\bibfnamefont {A.}~\bibnamefont {Maeda}}, \bibinfo {author} {\bibfnamefont
  {T.}~\bibnamefont {Takenobu}}, \bibinfo {author} {\bibfnamefont
  {Y.}~\bibnamefont {Iwasa}}, \ and\ \bibinfo {author} {\bibfnamefont
  {T.}~\bibnamefont {Mitani}},\ }\href@noop {} {\bibfield  {journal} {\bibinfo
  {journal} {Phys. Rev. Lett.}\ }\textbf {\bibinfo {volume} {88}},\ \bibinfo
  {pages} {096401} (\bibinfo {year} {2002})}\BibitemShut {NoStop}%
\bibitem [{\citenamefont {M\'arkus}\ \emph {et~al.}(2017)\citenamefont
  {M\'arkus}, \citenamefont {Simon}, \citenamefont {Nagy}, \citenamefont
  {Feh\'er}, \citenamefont {Wild}, \citenamefont {Abell\'an}, \citenamefont
  {Chac\'on‐Torres}, \citenamefont {Hirsch},\ and\ \citenamefont
  {Hauke}}]{MarkusPSSb2017}%
  \BibitemOpen
  \bibfield  {author} {\bibinfo {author} {\bibfnamefont {B.~G.}\ \bibnamefont
  {M\'arkus}}, \bibinfo {author} {\bibfnamefont {F.}~\bibnamefont {Simon}},
  \bibinfo {author} {\bibfnamefont {K.}~\bibnamefont {Nagy}}, \bibinfo {author}
  {\bibfnamefont {T.}~\bibnamefont {Feh\'er}}, \bibinfo {author} {\bibfnamefont
  {S.}~\bibnamefont {Wild}}, \bibinfo {author} {\bibfnamefont {G.}~\bibnamefont
  {Abell\'an}}, \bibinfo {author} {\bibfnamefont {J.~C.}\ \bibnamefont
  {Chac\'on‐Torres}}, \bibinfo {author} {\bibfnamefont {A.}~\bibnamefont
  {Hirsch}}, \ and\ \bibinfo {author} {\bibfnamefont {F.}~\bibnamefont
  {Hauke}},\ }\href@noop {} {\bibfield  {journal} {\bibinfo  {journal} {Phys.
  Status Solidi B}\ }\textbf {\bibinfo {volume} {254}},\ \bibinfo {pages}
  {1700232} (\bibinfo {year} {2017})}\BibitemShut {NoStop}%
\bibitem [{\citenamefont {Hone}\ \emph {et~al.}(1999)\citenamefont {Hone},
  \citenamefont {Whitney}, \citenamefont {Piskoti},\ and\ \citenamefont
  {Zettl}}]{HonePRB1999}%
  \BibitemOpen
  \bibfield  {author} {\bibinfo {author} {\bibfnamefont {J.}~\bibnamefont
  {Hone}}, \bibinfo {author} {\bibfnamefont {M.}~\bibnamefont {Whitney}},
  \bibinfo {author} {\bibfnamefont {C.}~\bibnamefont {Piskoti}}, \ and\
  \bibinfo {author} {\bibfnamefont {A.}~\bibnamefont {Zettl}},\ }\href@noop {}
  {\bibfield  {journal} {\bibinfo  {journal} {Phys. Rev. B}\ }\textbf {\bibinfo
  {volume} {59}},\ \bibinfo {pages} {R2514} (\bibinfo {year}
  {1999})}\BibitemShut {NoStop}%
\bibitem [{\citenamefont {Corzilius}\ \emph {et~al.}(2007)\citenamefont
  {Corzilius}, \citenamefont {Dinse}, \citenamefont {van Slageren},\ and\
  \citenamefont {Hata}}]{CorziliusPRB2007}%
  \BibitemOpen
  \bibfield  {author} {\bibinfo {author} {\bibfnamefont {B.}~\bibnamefont
  {Corzilius}}, \bibinfo {author} {\bibfnamefont {K.-P.}\ \bibnamefont
  {Dinse}}, \bibinfo {author} {\bibfnamefont {J.}~\bibnamefont {van Slageren}},
  \ and\ \bibinfo {author} {\bibfnamefont {K.}~\bibnamefont {Hata}},\
  }\href@noop {} {\bibfield  {journal} {\bibinfo  {journal} {Phys. Rev. B}\
  }\textbf {\bibinfo {volume} {75}},\ \bibinfo {pages} {235416} (\bibinfo
  {year} {2007})}\BibitemShut {NoStop}%
\bibitem [{\citenamefont {Corzilius}\ \emph {et~al.}(2008)\citenamefont
  {Corzilius}, \citenamefont {Dinse}, \citenamefont {Hata}, \citenamefont
  {Halu\v{s}ka}, \citenamefont {Sk\'{a}kalov\'{a}},\ and\ \citenamefont
  {Roth}}]{CorziliusPSSB2008}%
  \BibitemOpen
  \bibfield  {author} {\bibinfo {author} {\bibfnamefont {B.}~\bibnamefont
  {Corzilius}}, \bibinfo {author} {\bibfnamefont {K.-P.}\ \bibnamefont
  {Dinse}}, \bibinfo {author} {\bibfnamefont {K.}~\bibnamefont {Hata}},
  \bibinfo {author} {\bibfnamefont {M.}~\bibnamefont {Halu\v{s}ka}}, \bibinfo
  {author} {\bibfnamefont {V.}~\bibnamefont {Sk\'{a}kalov\'{a}}}, \ and\
  \bibinfo {author} {\bibfnamefont {S.}~\bibnamefont {Roth}},\ }\href@noop {}
  {\bibfield  {journal} {\bibinfo  {journal} {Phys. Status Solidi B}\ }\textbf
  {\bibinfo {volume} {245}},\ \bibinfo {pages} {2251} (\bibinfo {year}
  {2008})}\BibitemShut {NoStop}%
\bibitem [{\citenamefont {Karsa}\ \emph {et~al.}(2012)\citenamefont {Karsa},
  \citenamefont {Quintavalle}, \citenamefont {Forr\'o},\ and\ \citenamefont
  {Simon}}]{KarsaPSSB}%
  \BibitemOpen
  \bibfield  {author} {\bibinfo {author} {\bibfnamefont {A.}~\bibnamefont
  {Karsa}}, \bibinfo {author} {\bibfnamefont {D.}~\bibnamefont {Quintavalle}},
  \bibinfo {author} {\bibfnamefont {L.}~\bibnamefont {Forr\'o}}, \ and\
  \bibinfo {author} {\bibfnamefont {F.}~\bibnamefont {Simon}},\ }\href@noop {}
  {\bibfield  {journal} {\bibinfo  {journal} {Phys. Status Solidi B}\ }\textbf
  {\bibinfo {volume} {249}},\ \bibinfo {pages} {2487} (\bibinfo {year}
  {2012})}\BibitemShut {NoStop}%
\bibitem [{\citenamefont {Gy\"{u}re}\ \emph {et~al.}(2015)\citenamefont
  {Gy\"{u}re}, \citenamefont {G.~M\'arkus}, \citenamefont {Bern\'ath},
  \citenamefont {Mur\'anyi},\ and\ \citenamefont {Simon}}]{GyureRSI}%
  \BibitemOpen
  \bibfield  {author} {\bibinfo {author} {\bibfnamefont {B.}~\bibnamefont
  {Gy\"{u}re}}, \bibinfo {author} {\bibfnamefont {B.}~\bibnamefont
  {G.~M\'arkus}}, \bibinfo {author} {\bibfnamefont {B.}~\bibnamefont
  {Bern\'ath}}, \bibinfo {author} {\bibfnamefont {F.}~\bibnamefont
  {Mur\'anyi}}, \ and\ \bibinfo {author} {\bibfnamefont {F.}~\bibnamefont
  {Simon}},\ }\href@noop {} {\bibfield  {journal} {\bibinfo  {journal} {Review
  of Scientific Instruments}\ }\textbf {\bibinfo {volume} {86}},\ \bibinfo
  {pages} {094702} (\bibinfo {year} {2015})}\BibitemShut {NoStop}%
\bibitem [{\citenamefont {Gy\"{u}re-Garami}\ \emph {et~al.}()\citenamefont
  {Gy\"{u}re-Garami}, \citenamefont {S\'agi}, \citenamefont {M\'arkus},\ and\
  \citenamefont {Simon}}]{GyurePreprint}%
  \BibitemOpen
  \bibfield  {author} {\bibinfo {author} {\bibfnamefont {B.}~\bibnamefont
  {Gy\"{u}re-Garami}}, \bibinfo {author} {\bibfnamefont {O.}~\bibnamefont
  {S\'agi}}, \bibinfo {author} {\bibfnamefont {B.~G.}\ \bibnamefont
  {M\'arkus}}, \ and\ \bibinfo {author} {\bibfnamefont {F.}~\bibnamefont
  {Simon}},\ }\href@noop {} {\enquote {\bibinfo {title} {{A highly accurate
  measurement of resonator Q-factor and resonance frequency}},}\ }\bibinfo
  {howpublished} {arXiv:1805.11347}\BibitemShut {NoStop}%
\bibitem [{\citenamefont {Simon}\ \emph
  {et~al.}(2005{\natexlab{a}})\citenamefont {Simon}, \citenamefont {Kukovecz},
  \citenamefont {Kramberger}, \citenamefont {Pfeiffer}, \citenamefont {Hasi},
  \citenamefont {Kuzmany},\ and\ \citenamefont {Kataura}}]{SimonPRB2005}%
  \BibitemOpen
  \bibfield  {author} {\bibinfo {author} {\bibfnamefont {F.}~\bibnamefont
  {Simon}}, \bibinfo {author} {\bibfnamefont {{\'{A}}.}~\bibnamefont
  {Kukovecz}}, \bibinfo {author} {\bibfnamefont {C.}~\bibnamefont
  {Kramberger}}, \bibinfo {author} {\bibfnamefont {R.}~\bibnamefont
  {Pfeiffer}}, \bibinfo {author} {\bibfnamefont {F.}~\bibnamefont {Hasi}},
  \bibinfo {author} {\bibfnamefont {H.}~\bibnamefont {Kuzmany}}, \ and\
  \bibinfo {author} {\bibfnamefont {H.}~\bibnamefont {Kataura}},\ }\href@noop
  {} {\bibfield  {journal} {\bibinfo  {journal} {Phys. Rev. B}\ }\textbf
  {\bibinfo {volume} {71}},\ \bibinfo {pages} {165439} (\bibinfo {year}
  {2005}{\natexlab{a}})}\BibitemShut {NoStop}%
\bibitem [{\citenamefont {Simon}\ \emph
  {et~al.}(2006{\natexlab{a}})\citenamefont {Simon}, \citenamefont {Pfeiffer},\
  and\ \citenamefont {Kuzmany}}]{SimonPRB2006}%
  \BibitemOpen
  \bibfield  {author} {\bibinfo {author} {\bibfnamefont {F.}~\bibnamefont
  {Simon}}, \bibinfo {author} {\bibfnamefont {R.}~\bibnamefont {Pfeiffer}}, \
  and\ \bibinfo {author} {\bibfnamefont {H.}~\bibnamefont {Kuzmany}},\
  }\href@noop {} {\bibfield  {journal} {\bibinfo  {journal} {Phys. Rev. B}\
  }\textbf {\bibinfo {volume} {74}},\ \bibinfo {pages} {212411(R)} (\bibinfo
  {year} {2006}{\natexlab{a}})}\BibitemShut {NoStop}%
\bibitem [{\citenamefont {Simon}\ \emph
  {et~al.}(2005{\natexlab{b}})\citenamefont {Simon}, \citenamefont
  {Kramberger}, \citenamefont {Pfeiffer}, \citenamefont {Kuzmany},
  \citenamefont {Z\'{o}lyomi}, \citenamefont {K\"{u}rti}, \citenamefont
  {Singer},\ and\ \citenamefont {Alloul}}]{SimonPRL2005}%
  \BibitemOpen
  \bibfield  {author} {\bibinfo {author} {\bibfnamefont {F.}~\bibnamefont
  {Simon}}, \bibinfo {author} {\bibfnamefont {C.}~\bibnamefont {Kramberger}},
  \bibinfo {author} {\bibfnamefont {R.}~\bibnamefont {Pfeiffer}}, \bibinfo
  {author} {\bibfnamefont {H.}~\bibnamefont {Kuzmany}}, \bibinfo {author}
  {\bibfnamefont {V.}~\bibnamefont {Z\'{o}lyomi}}, \bibinfo {author}
  {\bibfnamefont {J.}~\bibnamefont {K\"{u}rti}}, \bibinfo {author}
  {\bibfnamefont {P.~M.}\ \bibnamefont {Singer}}, \ and\ \bibinfo {author}
  {\bibfnamefont {H.}~\bibnamefont {Alloul}},\ }\href@noop {} {\bibfield
  {journal} {\bibinfo  {journal} {Phys. Rev. Lett.}\ }\textbf {\bibinfo
  {volume} {95}},\ \bibinfo {pages} {017401} (\bibinfo {year}
  {2005}{\natexlab{b}})}\BibitemShut {NoStop}%
\bibitem [{\citenamefont {Singer}\ \emph {et~al.}(2005)\citenamefont {Singer},
  \citenamefont {Wzietek}, \citenamefont {Alloul}, \citenamefont {Simon},\ and\
  \citenamefont {Kuzmany}}]{SingerPRL2005}%
  \BibitemOpen
  \bibfield  {author} {\bibinfo {author} {\bibfnamefont {P.~M.}\ \bibnamefont
  {Singer}}, \bibinfo {author} {\bibfnamefont {P.}~\bibnamefont {Wzietek}},
  \bibinfo {author} {\bibfnamefont {H.}~\bibnamefont {Alloul}}, \bibinfo
  {author} {\bibfnamefont {F.}~\bibnamefont {Simon}}, \ and\ \bibinfo {author}
  {\bibfnamefont {H.}~\bibnamefont {Kuzmany}},\ }\href@noop {} {\bibfield
  {journal} {\bibinfo  {journal} {Phys. Rev. Lett.}\ }\textbf {\bibinfo
  {volume} {95}},\ \bibinfo {pages} {236403} (\bibinfo {year}
  {2005})}\BibitemShut {NoStop}%
\bibitem [{\citenamefont {Simon}\ \emph
  {et~al.}(2006{\natexlab{b}})\citenamefont {Simon}, \citenamefont {Kuzmany},
  \citenamefont {N\'{a}fr\'{a}di}, \citenamefont {Feh\'{e}r}, \citenamefont
  {Forr\'{o}}, \citenamefont {F\"{u}l\"{o}p}, \citenamefont {J\'{a}nossy},
  \citenamefont {Korecz}, \citenamefont {Rockenbauer}, \citenamefont {Hauke},\
  and\ \citenamefont {Hirsch}}]{SimonPRL2006}%
  \BibitemOpen
  \bibfield  {author} {\bibinfo {author} {\bibfnamefont {F.}~\bibnamefont
  {Simon}}, \bibinfo {author} {\bibfnamefont {H.}~\bibnamefont {Kuzmany}},
  \bibinfo {author} {\bibfnamefont {B.}~\bibnamefont {N\'{a}fr\'{a}di}},
  \bibinfo {author} {\bibfnamefont {T.}~\bibnamefont {Feh\'{e}r}}, \bibinfo
  {author} {\bibfnamefont {L.}~\bibnamefont {Forr\'{o}}}, \bibinfo {author}
  {\bibfnamefont {F.}~\bibnamefont {F\"{u}l\"{o}p}}, \bibinfo {author}
  {\bibfnamefont {A.}~\bibnamefont {J\'{a}nossy}}, \bibinfo {author}
  {\bibfnamefont {L.}~\bibnamefont {Korecz}}, \bibinfo {author} {\bibfnamefont
  {A.}~\bibnamefont {Rockenbauer}}, \bibinfo {author} {\bibfnamefont
  {F.}~\bibnamefont {Hauke}}, \ and\ \bibinfo {author} {\bibfnamefont
  {A.}~\bibnamefont {Hirsch}},\ }\href@noop {} {\bibfield  {journal} {\bibinfo
  {journal} {Phys. Rev. Lett.}\ }\textbf {\bibinfo {volume} {97}},\ \bibinfo
  {pages} {136801} (\bibinfo {year} {2006}{\natexlab{b}})}\BibitemShut
  {NoStop}%
\bibitem [{\citenamefont {Simon}\ \emph {et~al.}(2004)\citenamefont {Simon},
  \citenamefont {Kuzmany}, \citenamefont {Rauf}, \citenamefont {Pichler},
  \citenamefont {Bernardi}, \citenamefont {Peterlik}, \citenamefont {Korecz},
  \citenamefont {F\"ul\"op},\ and\ \citenamefont {J\'anossy}}]{SimonCPL2004}%
  \BibitemOpen
  \bibfield  {author} {\bibinfo {author} {\bibfnamefont {F.}~\bibnamefont
  {Simon}}, \bibinfo {author} {\bibfnamefont {H.}~\bibnamefont {Kuzmany}},
  \bibinfo {author} {\bibfnamefont {H.}~\bibnamefont {Rauf}}, \bibinfo {author}
  {\bibfnamefont {T.}~\bibnamefont {Pichler}}, \bibinfo {author} {\bibfnamefont
  {J.}~\bibnamefont {Bernardi}}, \bibinfo {author} {\bibfnamefont
  {H.}~\bibnamefont {Peterlik}}, \bibinfo {author} {\bibfnamefont
  {L.}~\bibnamefont {Korecz}}, \bibinfo {author} {\bibfnamefont
  {F.}~\bibnamefont {F\"ul\"op}}, \ and\ \bibinfo {author} {\bibfnamefont
  {A.}~\bibnamefont {J\'anossy}},\ }\href@noop {} {\bibfield  {journal}
  {\bibinfo  {journal} {Chem. Phys. Lett.}\ }\textbf {\bibinfo {volume}
  {383}},\ \bibinfo {pages} {362} (\bibinfo {year} {2004})}\BibitemShut
  {NoStop}%
\bibitem [{\citenamefont {Poole}(1983)}]{PooleBook}%
  \BibitemOpen
  \bibfield  {author} {\bibinfo {author} {\bibfnamefont {C.~P.}\ \bibnamefont
  {Poole}},\ }\href@noop {} {\emph {\bibinfo {title} {Electron Spin
  Resonance}}}\ (\bibinfo  {publisher} {John Wiley \& Sons},\ \bibinfo
  {address} {New York},\ \bibinfo {year} {1983})\BibitemShut {NoStop}%
\bibitem [{\citenamefont {Luiten}(2005)}]{LuitenReview}%
  \BibitemOpen
  \bibfield  {author} {\bibinfo {author} {\bibfnamefont {A.}~\bibnamefont
  {Luiten}},\ }\enquote {\bibinfo {title} {Q-factor measurements},}\ \
  (\bibinfo  {publisher} {John Wiley \& Sons, Inc.},\ \bibinfo {year}
  {2005})\BibitemShut {NoStop}%
\bibitem [{\citenamefont {Kajfez}(2005)}]{KajfezReview}%
  \BibitemOpen
  \bibfield  {author} {\bibinfo {author} {\bibfnamefont {D.}~\bibnamefont
  {Kajfez}},\ }\enquote {\bibinfo {title} {Q-factor},}\ \ (\bibinfo
  {publisher} {John Wiley \& Sons, Inc.},\ \bibinfo {year} {2005})\BibitemShut
  {NoStop}%
\bibitem [{\citenamefont {Cs\H{o}sz}\ \emph {et~al.}(2018)\citenamefont
  {Cs\H{o}sz}, \citenamefont {M\'arkus}, \citenamefont {J\'anossy},
  \citenamefont {Nemes}, \citenamefont {Mur\'anyi}, \citenamefont {Klupp},
  \citenamefont {Kamar\'as}, \citenamefont {Kogan}, \citenamefont {Bud'ko},
  \citenamefont {Canfield},\ and\ \citenamefont {Simon}}]{Csosz2018}%
  \BibitemOpen
  \bibfield  {author} {\bibinfo {author} {\bibfnamefont {G.}~\bibnamefont
  {Cs\H{o}sz}}, \bibinfo {author} {\bibfnamefont {B.~G.}\ \bibnamefont
  {M\'arkus}}, \bibinfo {author} {\bibfnamefont {A.}~\bibnamefont {J\'anossy}},
  \bibinfo {author} {\bibfnamefont {N.~M.}\ \bibnamefont {Nemes}}, \bibinfo
  {author} {\bibfnamefont {F.}~\bibnamefont {Mur\'anyi}}, \bibinfo {author}
  {\bibfnamefont {G.}~\bibnamefont {Klupp}}, \bibinfo {author} {\bibfnamefont
  {K.}~\bibnamefont {Kamar\'as}}, \bibinfo {author} {\bibfnamefont {V.~G.}\
  \bibnamefont {Kogan}}, \bibinfo {author} {\bibfnamefont {S.~L.}\ \bibnamefont
  {Bud'ko}}, \bibinfo {author} {\bibfnamefont {P.~C.}\ \bibnamefont
  {Canfield}}, \ and\ \bibinfo {author} {\bibfnamefont {F.}~\bibnamefont
  {Simon}},\ }\href@noop {} {\bibfield  {journal} {\bibinfo  {journal}
  {Available on arXiv:1804.08123}\ } (\bibinfo {year} {2018})}\BibitemShut
  {NoStop}%
\bibitem [{\citenamefont {Che}\ \emph {et~al.}(2000)\citenamefont {Che},
  \citenamefont {\c{C}a\u{g}{\i}n},\ and\ \citenamefont {{Goddard
  III}}}]{Che2000}%
  \BibitemOpen
  \bibfield  {author} {\bibinfo {author} {\bibfnamefont {J.}~\bibnamefont
  {Che}}, \bibinfo {author} {\bibfnamefont {T.}~\bibnamefont
  {\c{C}a\u{g}{\i}n}}, \ and\ \bibinfo {author} {\bibfnamefont {W.~A.}\
  \bibnamefont {{Goddard III}}},\ }\href@noop {} {\bibfield  {journal}
  {\bibinfo  {journal} {Nanotechnology}\ }\textbf {\bibinfo {volume} {11}},\
  \bibinfo {pages} {65} (\bibinfo {year} {2000})}\BibitemShut {NoStop}%
\bibitem [{\citenamefont {Grujicic}\ \emph {et~al.}(2004)\citenamefont
  {Grujicic}, \citenamefont {Cao},\ and\ \citenamefont
  {Gersten}}]{Grujicic2004}%
  \BibitemOpen
  \bibfield  {author} {\bibinfo {author} {\bibfnamefont {M.}~\bibnamefont
  {Grujicic}}, \bibinfo {author} {\bibfnamefont {G.}~\bibnamefont {Cao}}, \
  and\ \bibinfo {author} {\bibfnamefont {B.}~\bibnamefont {Gersten}},\
  }\href@noop {} {\bibfield  {journal} {\bibinfo  {journal} {Materials Science
  and Engineering}\ }\textbf {\bibinfo {volume} {B107}},\ \bibinfo {pages}
  {204} (\bibinfo {year} {2004})}\BibitemShut {NoStop}%
\bibitem [{\citenamefont {Han}\ and\ \citenamefont {Fina}(2011)}]{Han2011}%
  \BibitemOpen
  \bibfield  {author} {\bibinfo {author} {\bibfnamefont {Z.}~\bibnamefont
  {Han}}\ and\ \bibinfo {author} {\bibfnamefont {A.}~\bibnamefont {Fina}},\
  }\href@noop {} {\bibfield  {journal} {\bibinfo  {journal} {Progress in
  Polymer Science}\ }\textbf {\bibinfo {volume} {36}},\ \bibinfo {pages} {914}
  (\bibinfo {year} {2011})}\BibitemShut {NoStop}%
\bibitem [{\citenamefont {Yu}\ \emph {et~al.}(2005)\citenamefont {Yu},
  \citenamefont {Shi}, \citenamefont {Yao}, \citenamefont {Li},\ and\
  \citenamefont {Majumdar}}]{YuNanoLett2005}%
  \BibitemOpen
  \bibfield  {author} {\bibinfo {author} {\bibfnamefont {C.}~\bibnamefont
  {Yu}}, \bibinfo {author} {\bibfnamefont {L.}~\bibnamefont {Shi}}, \bibinfo
  {author} {\bibfnamefont {Z.}~\bibnamefont {Yao}}, \bibinfo {author}
  {\bibfnamefont {D.}~\bibnamefont {Li}}, \ and\ \bibinfo {author}
  {\bibfnamefont {A.}~\bibnamefont {Majumdar}},\ }\href@noop {} {\bibfield
  {journal} {\bibinfo  {journal} {Nano Letters}\ }\textbf {\bibinfo {volume}
  {5}},\ \bibinfo {pages} {1842} (\bibinfo {year} {2005})}\BibitemShut
  {NoStop}%
\bibitem [{\citenamefont {Hone}\ \emph {et~al.}(2000)\citenamefont {Hone},
  \citenamefont {Llaguno}, \citenamefont {Nemes}, \citenamefont {Johnson},
  \citenamefont {Fischer}, \citenamefont {Walters}, \citenamefont {Casavant},
  \citenamefont {Schmidt},\ and\ \citenamefont {Smalley}}]{Hone2000}%
  \BibitemOpen
  \bibfield  {author} {\bibinfo {author} {\bibfnamefont {J.}~\bibnamefont
  {Hone}}, \bibinfo {author} {\bibfnamefont {M.~C.}\ \bibnamefont {Llaguno}},
  \bibinfo {author} {\bibfnamefont {N.~M.}\ \bibnamefont {Nemes}}, \bibinfo
  {author} {\bibfnamefont {A.~T.}\ \bibnamefont {Johnson}}, \bibinfo {author}
  {\bibfnamefont {J.~E.}\ \bibnamefont {Fischer}}, \bibinfo {author}
  {\bibfnamefont {D.~A.}\ \bibnamefont {Walters}}, \bibinfo {author}
  {\bibfnamefont {M.~J.}\ \bibnamefont {Casavant}}, \bibinfo {author}
  {\bibfnamefont {J.}~\bibnamefont {Schmidt}}, \ and\ \bibinfo {author}
  {\bibfnamefont {R.~E.}\ \bibnamefont {Smalley}},\ }\href@noop {} {\bibfield
  {journal} {\bibinfo  {journal} {Applied Physics Letters}\ }\textbf {\bibinfo
  {volume} {77}},\ \bibinfo {pages} {666} (\bibinfo {year} {2000})}\BibitemShut
  {NoStop}%
\bibitem [{\citenamefont {Hone}\ \emph {et~al.}(2002)\citenamefont {Hone},
  \citenamefont {Llaguno}, \citenamefont {Biercuk}, \citenamefont {Johnson},
  \citenamefont {Batlogg}, \citenamefont {Benes},\ and\ \citenamefont
  {Fischer}}]{HoneAPA2002}%
  \BibitemOpen
  \bibfield  {author} {\bibinfo {author} {\bibfnamefont {J.}~\bibnamefont
  {Hone}}, \bibinfo {author} {\bibfnamefont {M.~C.}\ \bibnamefont {Llaguno}},
  \bibinfo {author} {\bibfnamefont {M.~J.}\ \bibnamefont {Biercuk}}, \bibinfo
  {author} {\bibfnamefont {A.~T.}\ \bibnamefont {Johnson}}, \bibinfo {author}
  {\bibfnamefont {B.}~\bibnamefont {Batlogg}}, \bibinfo {author} {\bibfnamefont
  {Z.}~\bibnamefont {Benes}}, \ and\ \bibinfo {author} {\bibfnamefont {J.~E.}\
  \bibnamefont {Fischer}},\ }\href@noop {} {\bibfield  {journal} {\bibinfo
  {journal} {Appl. Phys. A}\ }\textbf {\bibinfo {volume} {74}},\ \bibinfo
  {pages} {339} (\bibinfo {year} {2002})}\BibitemShut {NoStop}%
\bibitem [{\citenamefont {Fischer}\ \emph {et~al.}(2003)\citenamefont
  {Fischer}, \citenamefont {Zhou}, \citenamefont {Vavro}, \citenamefont
  {Llaguno}, \citenamefont {Guthy}, \citenamefont {Haggenmueller},
  \citenamefont {Casavant}, \citenamefont {Walters},\ and\ \citenamefont
  {Smalley}}]{Fischer2003}%
  \BibitemOpen
  \bibfield  {author} {\bibinfo {author} {\bibfnamefont {J.~E.}\ \bibnamefont
  {Fischer}}, \bibinfo {author} {\bibfnamefont {W.}~\bibnamefont {Zhou}},
  \bibinfo {author} {\bibfnamefont {J.}~\bibnamefont {Vavro}}, \bibinfo
  {author} {\bibfnamefont {M.~C.}\ \bibnamefont {Llaguno}}, \bibinfo {author}
  {\bibfnamefont {C.}~\bibnamefont {Guthy}}, \bibinfo {author} {\bibfnamefont
  {R.}~\bibnamefont {Haggenmueller}}, \bibinfo {author} {\bibfnamefont {M.~J.}\
  \bibnamefont {Casavant}}, \bibinfo {author} {\bibfnamefont {D.~E.}\
  \bibnamefont {Walters}}, \ and\ \bibinfo {author} {\bibfnamefont {R.~E.}\
  \bibnamefont {Smalley}},\ }\href@noop {} {\bibfield  {journal} {\bibinfo
  {journal} {Journal of Applied Physics}\ }\textbf {\bibinfo {volume} {96}},\
  \bibinfo {pages} {2157} (\bibinfo {year} {2003})}\BibitemShut {NoStop}%
\bibitem [{\citenamefont {Sinha}\ \emph {et~al.}(2005)\citenamefont {Sinha},
  \citenamefont {Barjami}, \citenamefont {Iannacchione}, \citenamefont
  {Schwab},\ and\ \citenamefont {Muench}}]{SinhaNanopart2005}%
  \BibitemOpen
  \bibfield  {author} {\bibinfo {author} {\bibfnamefont {S.}~\bibnamefont
  {Sinha}}, \bibinfo {author} {\bibfnamefont {S.}~\bibnamefont {Barjami}},
  \bibinfo {author} {\bibfnamefont {G.}~\bibnamefont {Iannacchione}}, \bibinfo
  {author} {\bibfnamefont {A.}~\bibnamefont {Schwab}}, \ and\ \bibinfo {author}
  {\bibfnamefont {G.}~\bibnamefont {Muench}},\ }\href@noop {} {\bibfield
  {journal} {\bibinfo  {journal} {Journal of Nanoparticle Research}\ }\textbf
  {\bibinfo {volume} {7}},\ \bibinfo {pages} {651} (\bibinfo {year}
  {2005})}\BibitemShut {NoStop}%
\end{thebibliography}%



\end{document}